\documentclass[traditabstract]{aa}  
\usepackage{graphicx}
\usepackage{txfonts}
\usepackage{rotating}
\usepackage{natbib}
\usepackage[usenames]{color}
\usepackage{multirow}
\usepackage{url}
\usepackage{ulem}
\bibpunct{(}{)}{;}{a}{}{,} 
\newcommand{\ratio} {N({\rm H}_2) / I_{\rm CO(1-0)}}
\newcommand{\ratioo} {N({\rm H}_2) / I_{\rm CO}}

\newcommand{\kms}   {{\rm \  km \  s^{-1}}}

\newcommand{\Xunit} {\,{\rm cm^{-2}/(K\kms)}}

\newcommand{\msun}{M$_{\odot}$}

\begin{document}

\title{ Properties and rotation of molecular clouds in M~33 }
		
\author{J. Braine\inst{1} \and E. Rosolowsky\inst{2}  \and P. Gratier\inst{1} \and E. Corbelli\inst{3}  \and K.-F. Schuster\inst{4} }

\institute{Laboratoire d'Astrophysique de Bordeaux, Univ. Bordeaux, CNRS, B18N, all\'ee Geoffroy Saint-Hilaire, 33615 Pessac, France.\\
             \email{jonathan.braine@u-bordeaux.fr}
        \and Department of Physics, University of Alberta, Edmonton, AB, T6G 2E1, Canada
        \and INAF-Osservatorio Astrofisico di Arcetri, Largo E. Fermi, 5, 50125 Firenze, Italy
        \and Institut de Radioastronomie Millim\'etrique, 300 rue de la Piscine, Domaine Universitaire 38406 Saint Martin d'H\`eres, France
        }
\date{Received xxxx; accepted xxxx}

\abstract {The sample of 566 molecular clouds identified in the CO(2--1) IRAM survey covering the disk of M~33 is explored in detail.  The clouds were found using CPROPS and were subsequently catalogued in terms of their star-forming properties as non-star-forming (A), with embedded star formation (B), or with exposed star formation (C, e.g. presence of H$\alpha$ emission).
We find that the size-linewidth relation among the M~33 clouds is quite weak but, when comparing with clouds in other nearby galaxies, the linewidth scales
with average metallicity.  The linewidth and particularly the line brightness decrease with galactocentric distance.  The large number of clouds makes it possible to calculate well-sampled cloud mass spectra and mass spectra of subsamples.  
As noted earlier, but considerably better defined here, the mass spectrum steepens (i.e. higher fraction of small clouds) with galactocentric distance.  A new finding is that the mass spectrum of A clouds is much steeper than that of the star-forming clouds.  Further dividing the sample, this difference is strong at both large and small galactocentric distances and the A vs C difference is a stronger effect than the inner/outer disk difference in mass spectra.
Velocity gradients are identified in the clouds using standard techniques.  The gradients are weak and are dominated by prograde rotation; the effect is stronger for the high signal-to-noise clouds.  A discussion of the uncertainties is presented. The angular momenta are low but compatible with at least some simulations.  Finally, the cloud velocity gradients are compared with the gradient of disk rotation.
The cloud and galactic gradients are similar; the cloud rotation periods are much longer than cloud lifetimes and comparable to the galactic rotation period.  The rotational kinetic energy is $1-2$\% of the gravitational potential energy and the cloud edge velocity is well below the escape velocity, such that cloud-scale rotation probably has little influence on the evolution of molecular clouds.  
  }	

\keywords{Galaxies: Individual: M~33 --  -- Galaxies: Local Group -- Galaxies: ISM -- ISM: clouds -- ISM: Molecules -- Stars:
    Formation  }

\maketitle

\section{Introduction}

Recent years have seen a sharply increasing number of studies of molecular clouds in external galaxies \citep{Wilson90, Engargiola03, Rosolowsky03, Kawamura09, Gratier10b, Gratier12, Donovan-Meyer13, Schinnerer2013, Druard14, Corbelli17, Freeman17}.  
All of these authors wanted to know whether we could apply knowledge of molecular clouds in the Milky Way to other 
galaxies and other environments.  Of particular interest was the metallicity but also the morphology of the host galaxy.

M~33 is a small spiral galaxy in the Local Group with a thin disk and a metallicity about half that of the solar neighborhood.  It is at about 840~kpc \citep{Galleti04} and rather ideally inclined with an inclination of $56^\circ$, such that cloud velocities are straightforwardly separated and the structure of the disk is plainly visible.  The full disk of M~33 was observed in the CO(2--1) line with the IRAM 30meter radiotelescope \citep[see details in][]{Gardan07,Gratier10,Druard14}, yielding a resolution of 12$"$ or 49~pc.
The CPROPS software \citet{Rosolowsky06} was used to extract a 566 cloud sample from the CO data cube \citep[see][for details]{Gratier12,Druardthesis} and published in \citet{Corbelli17}. 
The 566 clouds identified in M33 were classified in terms of their star formation by \citet{Corbelli17} \citep[following ][]{Gratier12} and here we use this classification to compare star-forming clouds from those with no observed star formation.
Figure 1 (left) shows the CO(2--1) integrated intensity map with the cloud contours superposed.  

We use the cloud sample to investigate trends in linewidths in M~33 and in the broader context of 
Local Group galaxies and M51.  We specifically look for a connection between galaxy metallicity and the 
cloud size-linewidth relation. Some of the work based on a partial cloud sample and presented in 
\citet{Gratier12} is redone with the full cloud sample.  The link between cloud mass spectra and star 
formation is examined and Herschel SPIRE photometry data are used to estimate dust temperatures 
for the star-forming and non-star-forming clouds.

In this work we study not only the general properties of the clouds but also look for velocity gradients across the clouds as signs of possible cloud rotation.  Assuming that the medium out of which clouds form is larger than the molecular cloud, angular momentum conservation should result in detectable rotation velocities \citep[see e.g.][]{Rosolowsky03}.  The first to measure cloud velocity gradients, in order to trace cloud rotation, were \citet{Kutner77}
who estimated a large-scale velocity gradient of  $0.135\kms {\rm pc}^{-1}$ opposite to galactic rotation (retrograde).  These authors argue that the velocity gradient is due to rotation.  In the review by \citet{Blitz93}, it is concluded that velocity gradients are probably the result of rotation but that individual clouds (they identify W3) could be exceptions.  A more detailed discussion of Milky Way cloud velocity gradients can be found in \citet{Phillips99}.  Our observations are GMC-scale, as opposed to much smaller scales where rotating disks are apparent, so influences other than rotation may be present in our measures (see Section 3). 

\citet{Rosolowsky03} were the first to look at velocity gradients in an external galaxy.  Extragalactic work is quite complementary to the Galactic observations as the biases are not at all the same although the spatial resolution is much poorer.  They identified 45 molecular clouds in M~33 and found velocity gradients to be very low, with nearly half in the retrograde direction.  \citet{Imara11b} looked at the sample and suggested that in fact the clouds may not be rotating.  We use our sample of 566 clouds to re-evaluate the question of cloud rotation. 


\begin{figure*}
	\centering
	\includegraphics[width=\hsize{}]{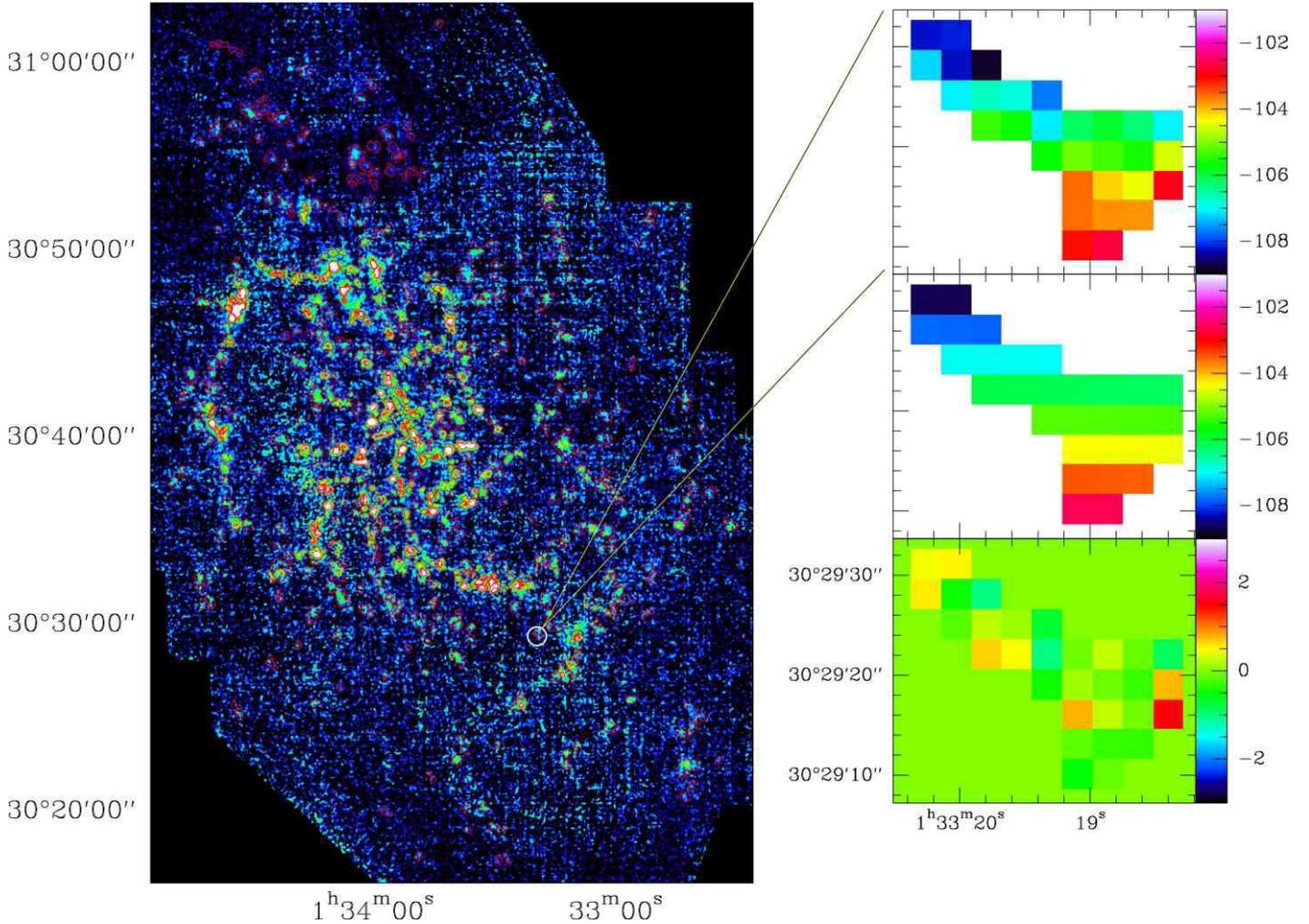}
	\caption{(left) CO(2--1) integrated intensity map of M33 with the cloud contours superposed in red and a zoom on cloud \#4 showing (top right) the velocities as measured using Eq (1) , (middle right) the fit to the velocities using Eq (2), and to the bottom right the velocity residual, all in $\kms$. }
	\label{rot_fig1} 
\end{figure*}


\section{Cloud properties}

In addition to estimating 3D cloud boundaries, CPROPS generates information such as deconvolved sizes and cloud luminosities.  Outside the CPROPS calculations, we estimate cloud linewidths as in \citet{Gratier12} by fitting a gaussian line profile to the average cloud profile.  Cloud masses are estimated by assuming a constant $\ratio$ factor of $4 \times 10^{20} \Xunit$ and a CO($\frac{2-1}{1-0}$) line ratio of 0.8.  These values have been validated by \citet{Druard14}, \citet{Gratier17}, and \citet{Braine10b}.  

\begin{figure}
	\centering
	\includegraphics[width=\hsize{}]{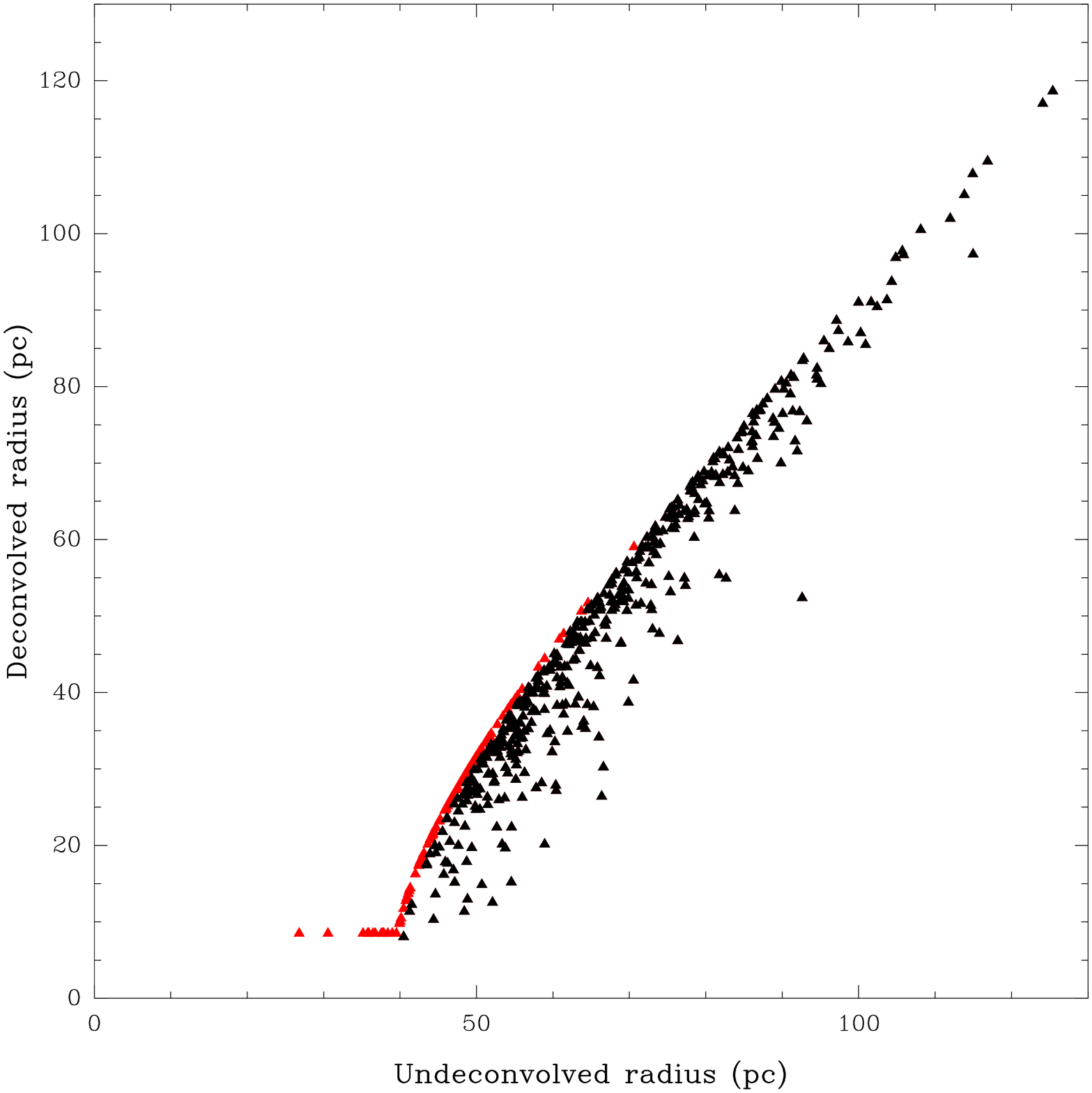}
	\caption{ Link between CPROPS undeconvolved cloud radii (x-axis) and deconvolved radii for the 80\% of the clouds with well-defined deconvolved  radii (black symbols).  The red triangles show the radii attributed to the clouds for which CPROPS could not estimate radii -- the radii follow the envelope defined by the CPROPS values down to a constant value below which we have no confidence.  The goal here is to avoid generating clouds which are inappropriately small while still attributing reasonable radii. }
	\label{decon} 
\end{figure}

CPROPS was able to estimate deconvolved radii for 449 out of 566 clouds.  In order to provide radii for the remaining 20\%, we use the link between non-deconvolved and deconvolved radii for the 449 clouds to extrapolate to determine deconvolved radii for the remaining clouds.  This enables the use of the whole sample but does not change the results.
Figure~\ref{decon}  shows the relation we fit with the 449 clouds (black) and the values calculated for the others (red symbols). 
From now on, we use the whole set of deconvolved cloud radii when radii are used.
Over the whole disk, about half the CO emission comes from identified clouds.  The fraction decreases with galactocentric distance: 2/3 of the CO emission is in clouds in the inner disk and 1/3 in the outer disk \citep{Druardthesis}.

\subsection{Size-linewidth relation}

\begin{figure}
	\centering
	\includegraphics[width=\hsize{}]{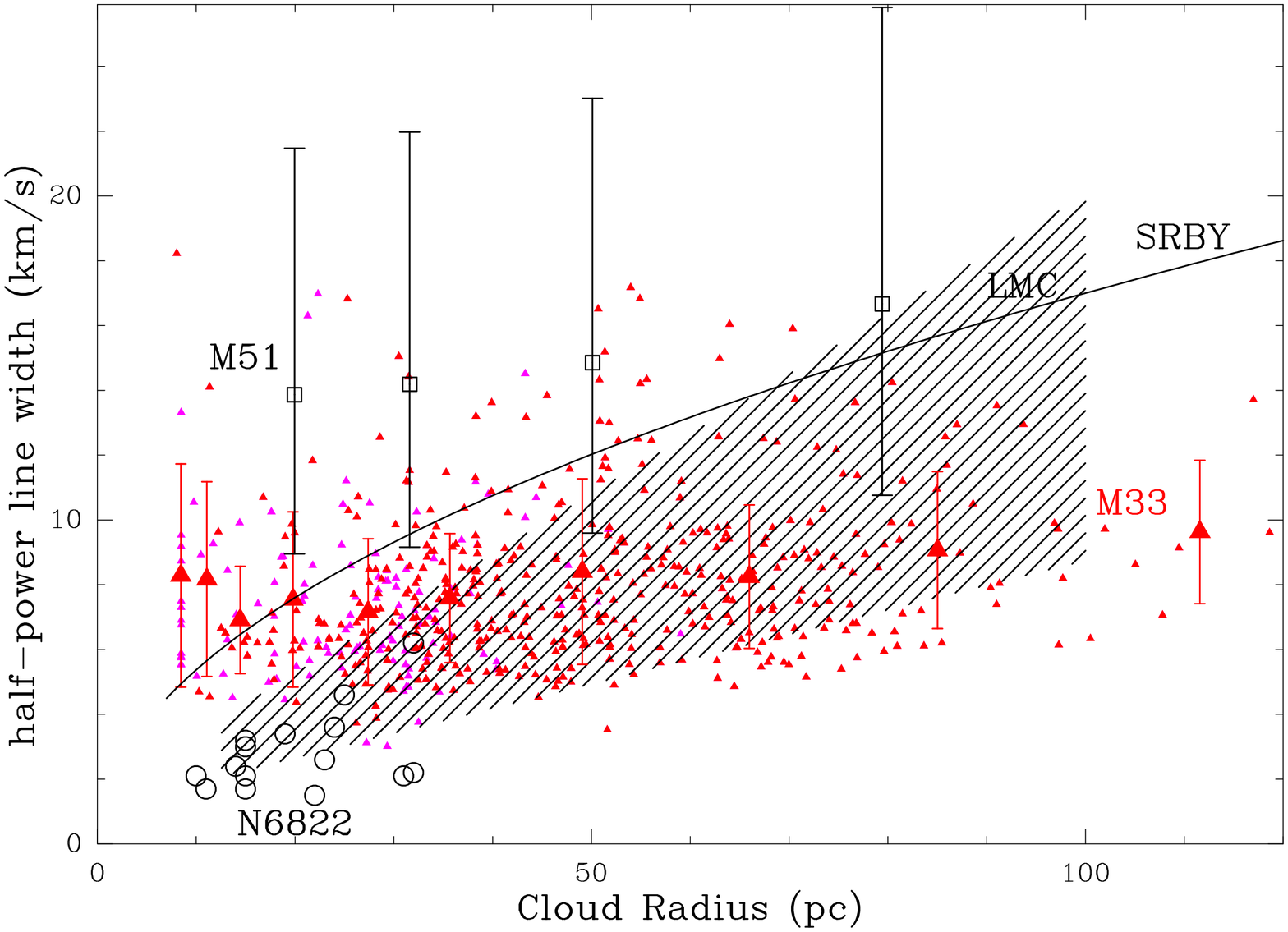}
	\includegraphics[width=\hsize{}]{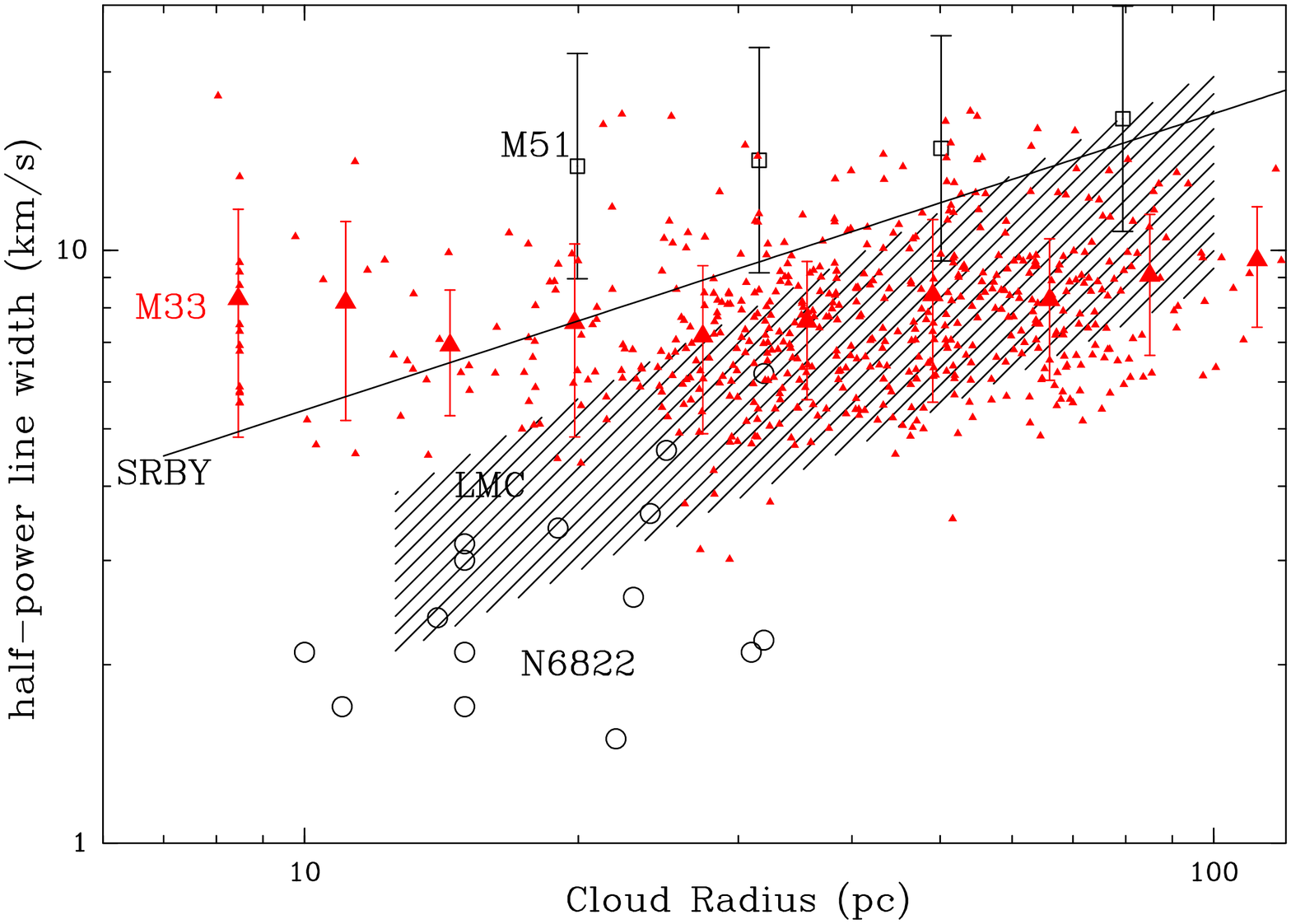}
	\caption{Size-linewidth relation for M33 clouds and other galaxies observed with similar or better angular resolution.  In the top panel, the clouds with extrapolated radii are shown as magenta symbols and the remaining 80\% of the M~33 clouds are shown in red. 
	The non-M~33 clouds are shown as black symbols. The region populated by LMC clouds is hatched and the M51 data (squares) are binned averages. The lower panel has logarithmic axes and has all M~33 clouds in red. The line labelled SRBY is the \citet{Solomon87} relation for the Galaxy. }
	\label{size-linewid} 
\end{figure}

Figure~\ref{size-linewid} shows the size-linewidth relation for the M33 clouds (red symbols) as compared to other nearby galaxies for which similar data are available, including our own Galaxy.  The \citet{Solomon87} relation for the Milky Way is shown as a line.  M~51 data come from \citet{Colombo14}, the LMC region from \citet{Hughes10}, and the NGC~6822 data from \citet{Gratier10b}.

Two things are apparent from Figure~\ref{size-linewid}.  First, the size-linewidth relation is very weak in M~33, NGC~6822, and M~51, but apparently strong in the Galaxy and the LMC, although Fig. 4a of \citet{Hughes10} shows that there is a very high dispersion in the relation in the LMC.  Second, the smaller (and lower metallicity) galaxies have clouds with narrower lines at a given size.  {\it To our knowledge, this is the first time this has been noticed.} 
Given the link between metallicity and galaxy size or mass, it is not clear whether the narrower lines are due to the change in metallicity or the change in stellar surface density.

Nonetheless, this shows that molecular clouds have distinctly different properties in different types of galaxies.
A number of studies have shown that, even after correcting for a varying $\ratio$ conversion factor, the molecular gas consumption time is lower in low-metallicity galaxies \citep{Gardan07, Gratier10, Braine10b, Dib11, Druard14}.  This is likely partially due to the weaker stellar winds, slowing cloud dispersal, in subsolar metallicity stars, but also because molecular gas is likely to form at slightly higher densities due to the reduced dust content.  Both factors could result in lower cloud line widths for a given size.  Clearly, a change in the stellar Initial Mass Function could greatly affect the elements of the above calculation.

\begin{figure}
	\centering
	\includegraphics[width=\hsize{}]{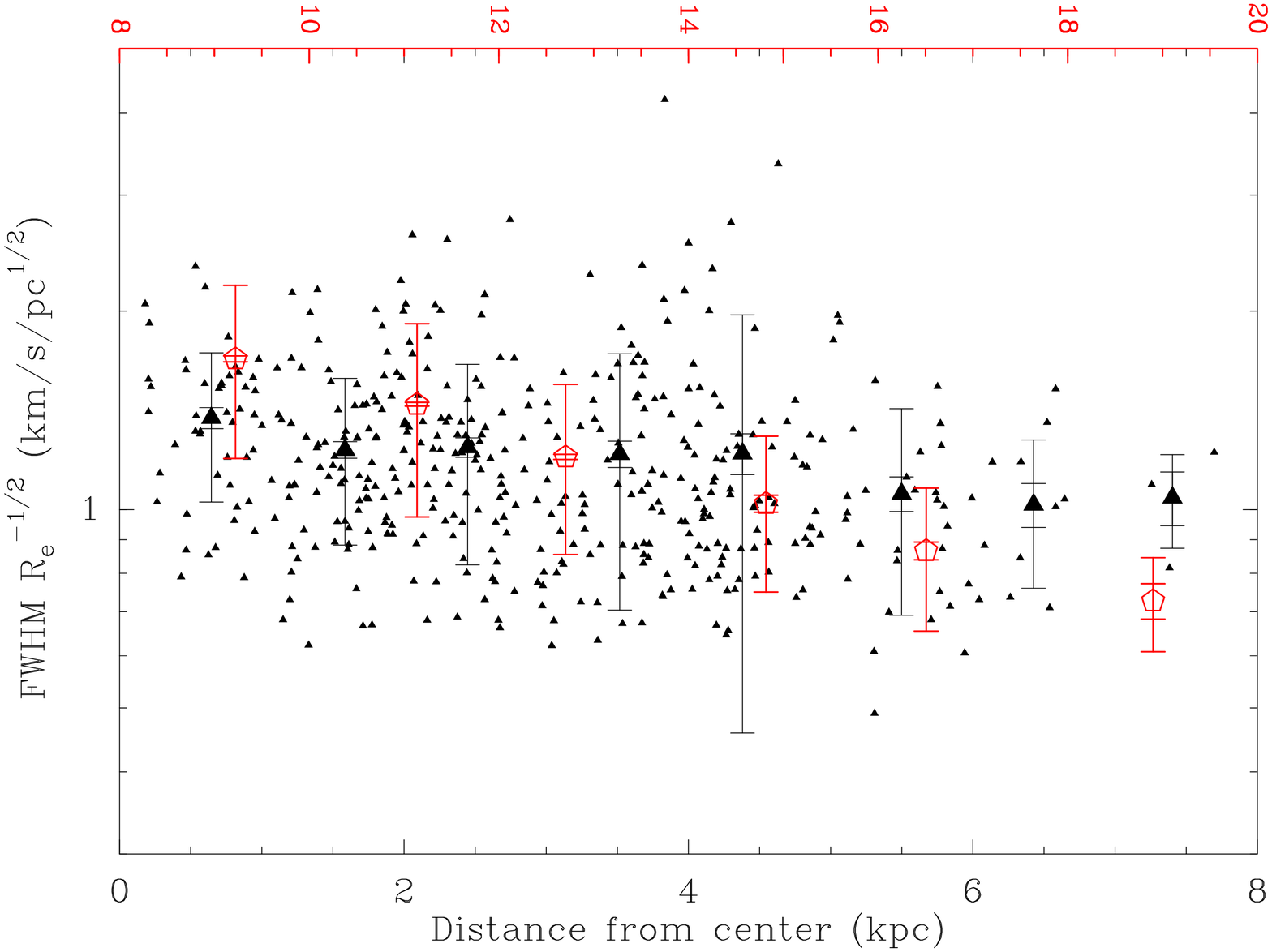}
	\caption{ Turbulent line width on a 1 pc scale as a function of galactocentric distance in M33 (black) and the outer Milky Way (red).  Individual M~33 clouds are shown as small black triangles and the large symbols are the binned averages for the M33 clouds (large black triangles) and for the \citet{Heyer01} outer Galaxy clouds (red pentagons and red upper distance scale). The large and small error bars on both symbols indicate respectively the dispersion and the uncertainty in the mean on the average values. Both galaxies show a modest decrease with galactocentric radius.}
	\label{erik3} 
\end{figure}

Another way of looking at the size-linewidth relation and its variation is shown in Fig.~\ref{erik3}.
Here we plot the linewidth normalized by the square root of the cloud radius, in order to eliminate the standard size-linewidth relation.
The \citet{Solomon87} relation would be a horizontal line at 1.7 $\kms$ pc$^{-1/2}$ in this plot.  
The M33 clouds are seen as small black triangles and they are binned by radial intervals (large black triangles) to show how this quantity, which can be thought of as the turbulent line width on a 1 pc scale, varies with galactocentric distance.  A clear decrease can be seen with distance from the center.  The large error bars indicate the dispersion and the small red error bars indicate the uncertainty in the mean value.

The radial distances in M~33 cover a range of about 7kpc, with few clouds beyond 6.5 kpc.  
All of the M~33 averages fall below the \citet{Solomon87} relation.  
For comparison, we took the large sample of \citet{Heyer01} for the outer disk of the Milky Way.  
Taking the clouds larger than 1~pc, we calculated the linewidth normalized by the square root of the cloud radius and binned by as a function of the distance from the Galactic center.  These points are shown as red pentagons and the distance is given on the red upper scale, covering nearly 12kpc from the solar circle to 20 kpc.  As for the black triangles, the large error bars give the dispersion and the small error bars give the uncertainty in the mean value.  The decrease in the turbulent line width on a 1 pc scale is very similar to that in M33, confirming that indeed cloud linewidths decrease with increasing galactocentric distance.

\subsection{Linewidth variation with galactocentric distance}

\begin{figure}
	\centering
	\includegraphics[width=\hsize{}]{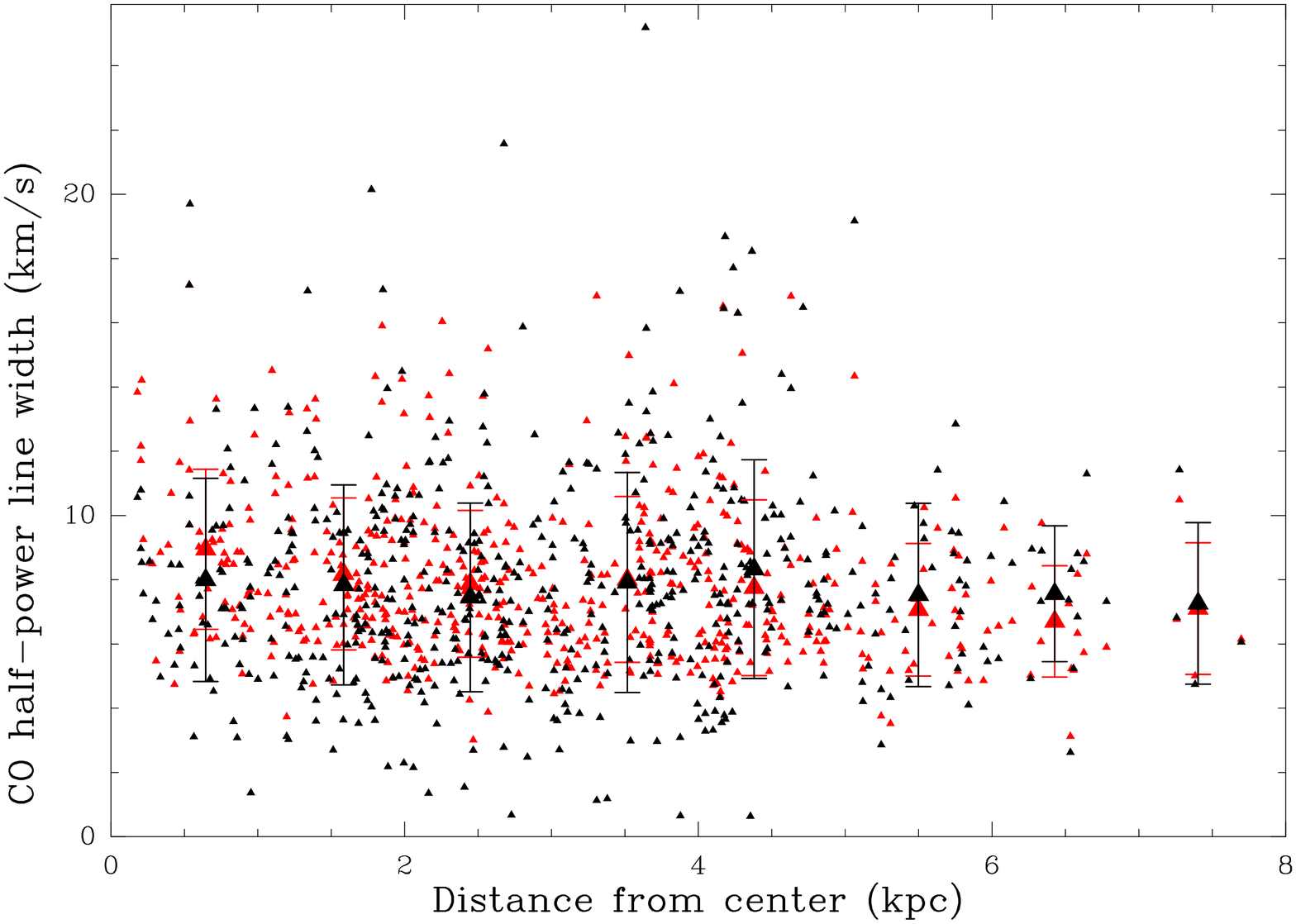}
	\caption{ Variation of the cloud linewidth versus galactocentric distance. Black symbols are the line widths from CPROPS and red symbols are from fitting gaussians to the cloud spectra.  Errorbars indicate the dispersion within each bin, showing that generally the gaussian fits yield fewer extreme results \citep[as found by ][]{Gratier12}.}
	\label{dist-dvco} 
\end{figure}

It is known that the HI line widths in galaxies decrease with galactocentric distance \citep{Tamburro09} but few measures are available for  molecular lines of individual clouds.  \citet{Braine10a} reported a decrease in their sample of a few clouds in M33 and \citet{Gratier12} reported a weak decline in linewidth.  Figure~\ref{dist-dvco} shows the results for the whole  sample of 566 clouds.  The decline in linewidth is only slight but is significant at the $8\sigma$ level, similar to what was found by \citet{Gratier12}. 
The uncertainty is obtained by bootstrapping, using 5000 iterations each choosing 566 clouds randomly out of the sample (allowing clouds to be chosen more than once or not at all) and examining the dispersion in the correlation coefficient of the fit between galactocentric distance and linewidth \citep[see details in ][]{Gratier12}.  It can be seen that the dispersion in line widths is generally lower for the gaussian fits so these were used in Figure~\ref{size-linewid}.

\subsection{ Cloud brightness as a function of galactocentric distance}

\begin{figure}
	\centering
	\includegraphics[width=\hsize{}]{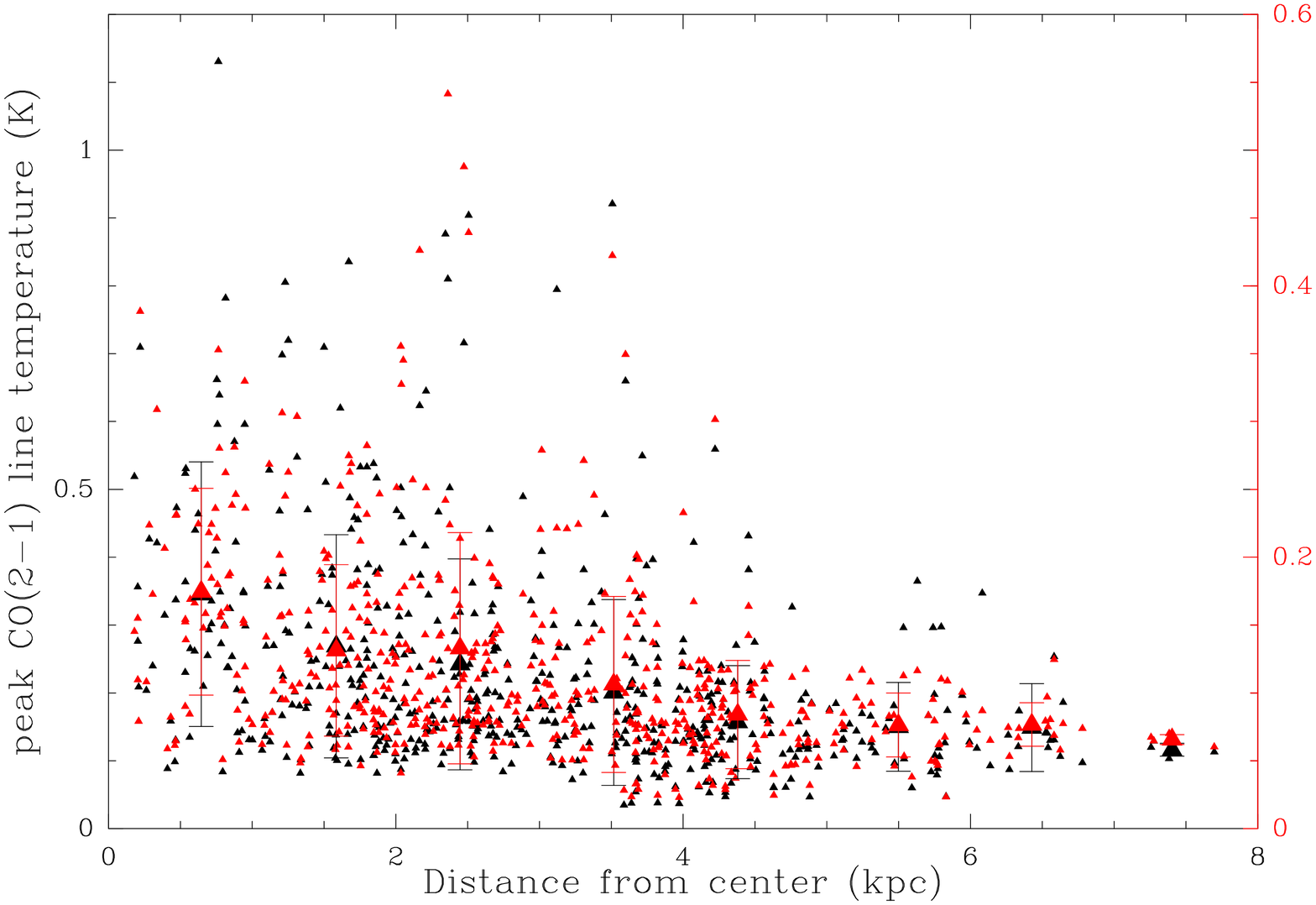}
	\caption{Galactocentric radius (x-axis) versus peak line temperature reached within the cloud (black, left y-scale)  and the peak of the gaussian fit to the line profile (red symbols, right scale).  The factor two difference between the left and right scales is because the peak line temperature is from a single position and is about twice as strong as the gaussian fit to the cloud-averaged line profile.  Both temperatures decrease in a similar way.}
	\label{dist-tco} 
\end{figure}

It is well-known that the large-scale CO brightness of galaxies decreases with distance from the center in most galaxies \citep[see][for survey and M33 results respectively]{Young95, Druard14}.  Notable counter-examples are M~31 and M81.  \citet{Gratier12} showed that for the then available cloud sample in M~33 the clouds became considerably less CO-bright with increasing distance to the center.  Here we use the entire cloud sample and two different measures of brightness:  Fig.~\ref{dist-tco} shows the peak line temperature reached within the cloud and the peak of the gaussian fit to the cloud-averaged line profile.  The decrease in both quantities is approximately a factor two over 4 kpc, corresponding to a scale length of about 6 kpc.  The average CO surface brightness per kpc$^2$ decreases much more quickly, with a scale length of approximately 2 kpc \citep{Druard14}.

\subsection{Mass spectrum of M~33 clouds}

The sample of molecular clouds analyzed here, whose individual cloud properties have been
given in the on-line table of \citet{Corbelli17} is one of the largest of any galaxy
and the largest for which a classification in terms of star formation has been established. 
We fit a truncated power law to the distribution of cloud masses above the completeness limit, following the procedure described by \citet{Maschberger09} and the methods described in \citet{Gratier12}, to determine mass spectra for the sample and subcategories, making the assumption as elsewhere that the CO luminosity reflects the mass of clouds \citep[cf.][]{Corbelli17}.  

\begin{figure}
	\centering
	\includegraphics[width=\hsize{}]{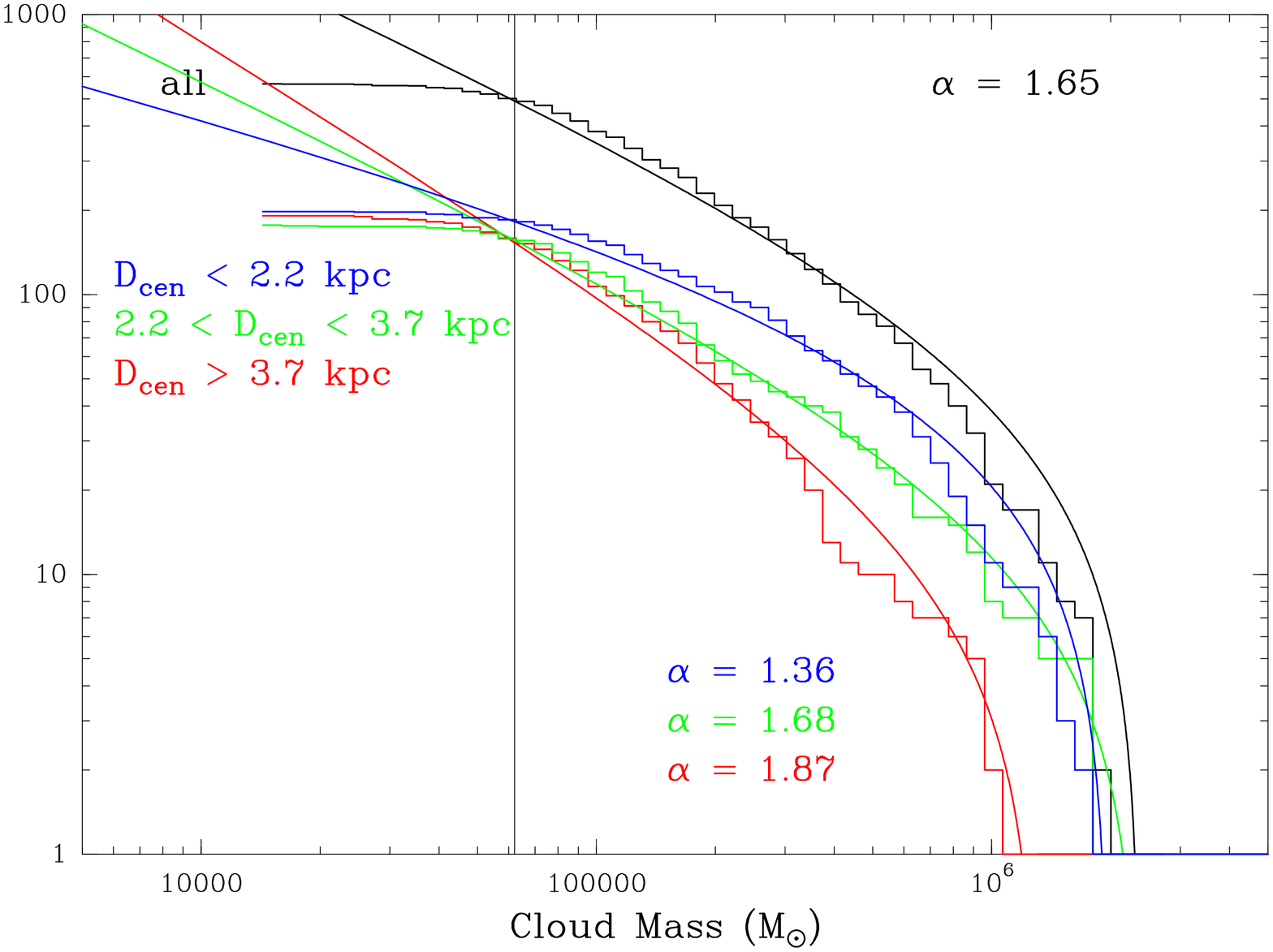}
	\includegraphics[width=\hsize{}]{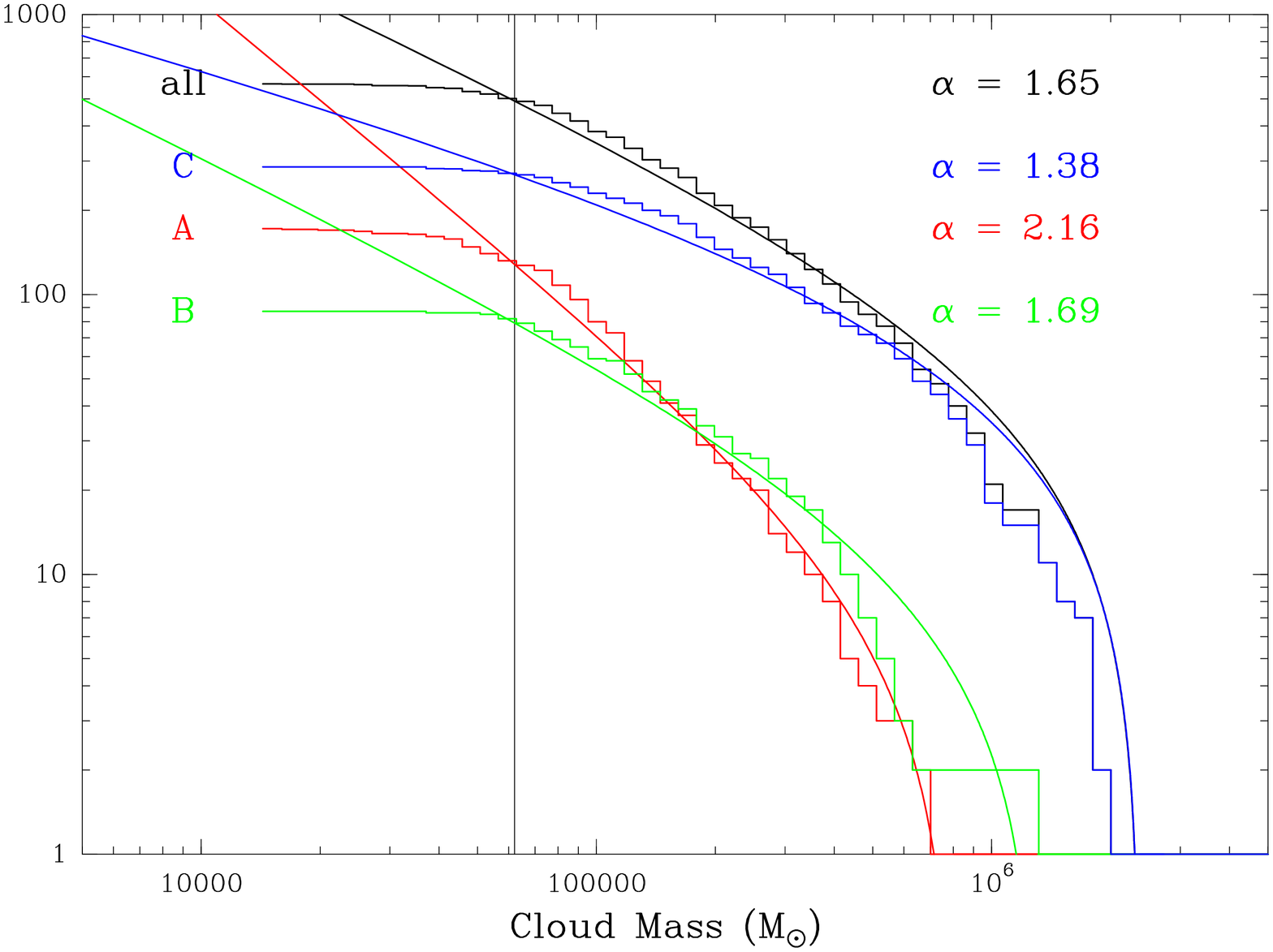}
	\caption{ Mass spectra of M~33 molecular clouds.  The y-axis gives the cumulative number of clouds with mass above the corresponding x-axis mass.  The vertical line shows the completeness limit as in \citet{Corbelli17}.  The black line shows the whole sample in both panels.  The color-coded $\alpha$ values give the slopes of the mass spectra.  In the top panel, clouds are segregated by galactocentric radius as indicated -- the sample has been divided into 3 roughly equally populated radial bins. In the lower panel, the types indicate clouds with exposed star formation ({\it e.g.} H$\alpha$ emission), embedded star formation, and no star formation, denoted respectively C, B, and A types. The division into types is discussed in \citet{Corbelli17} and \citet{Gratier12}. The solid lines are the results of the fits, with the color corresponding to the (sub)sample. }
	\label{mass_spec} 
\end{figure}

For the entire sample of 566 clouds, we obtain an exposant $\alpha=1.65$ defining the slope of the mass spectrum $n(m)dm \propto m^{-\alpha} dm$ (Fig~\ref{mass_spec}).  The upper panel shows, as seen previously by \citet{Gratier12} and \citet{Rosolowsky05} that the mass spectrum steepens in the outer disk.  
With the larger sample, we are able to divide the sample into 3 radial bins with close to 200 clouds per bin, far more than earlier samples.  The slope of the mass function steepens from $\alpha=1.4$ to  $\alpha=1.9$ with radius and would probably continue to steepen but the number of clouds available (above the completeness limit) decreases sharply, such that the slope becomes poorly defined.  The cause of such a steepening remains unclear so we also compared the mass spectra of the clouds at different stages in the star formation process.

The lower panel of Fig~\ref{mass_spec} compares the sub-sample of non-star-forming clouds (A) with clouds showing embedded and exposed star formation \citep[respectively B and C clouds, ][]{Corbelli17}.
The spectra are remarkably different: the C clouds are more massive and have a distinctly flatter spectrum than, particularly, the clouds without star formation.  The B clouds, less numerous, are intermediate.  This behavior mimics the radial variation and indeed there is some degeneracy as the star-forming clouds are on average closer to the center than those without star formation.  
While the cloud classification system is not the same, \citet{Kawamura09} also found that the more evolved clouds were more massive.  We \citep{Corbelli17} attribute this to continued gas accretion.
Since we measure CO emission, and not mass directly, a systematic variation in the $\ratioo$ conversion factor could generate a similar result.  There are two reasons for thinking that this is unlikely.  A sophisticated bayesian analysis of the $\ratioo$ factor (and "dark gas") by \citet{Gratier17} found no identifiable variation in the $\ratioo$ conversion as a function of radius in M~33.  Secondly, the dust-based cloud masses from the Herschel data show the same trends with star formation class.

\begin{figure}
	\centering
	\includegraphics[width=\hsize{}]{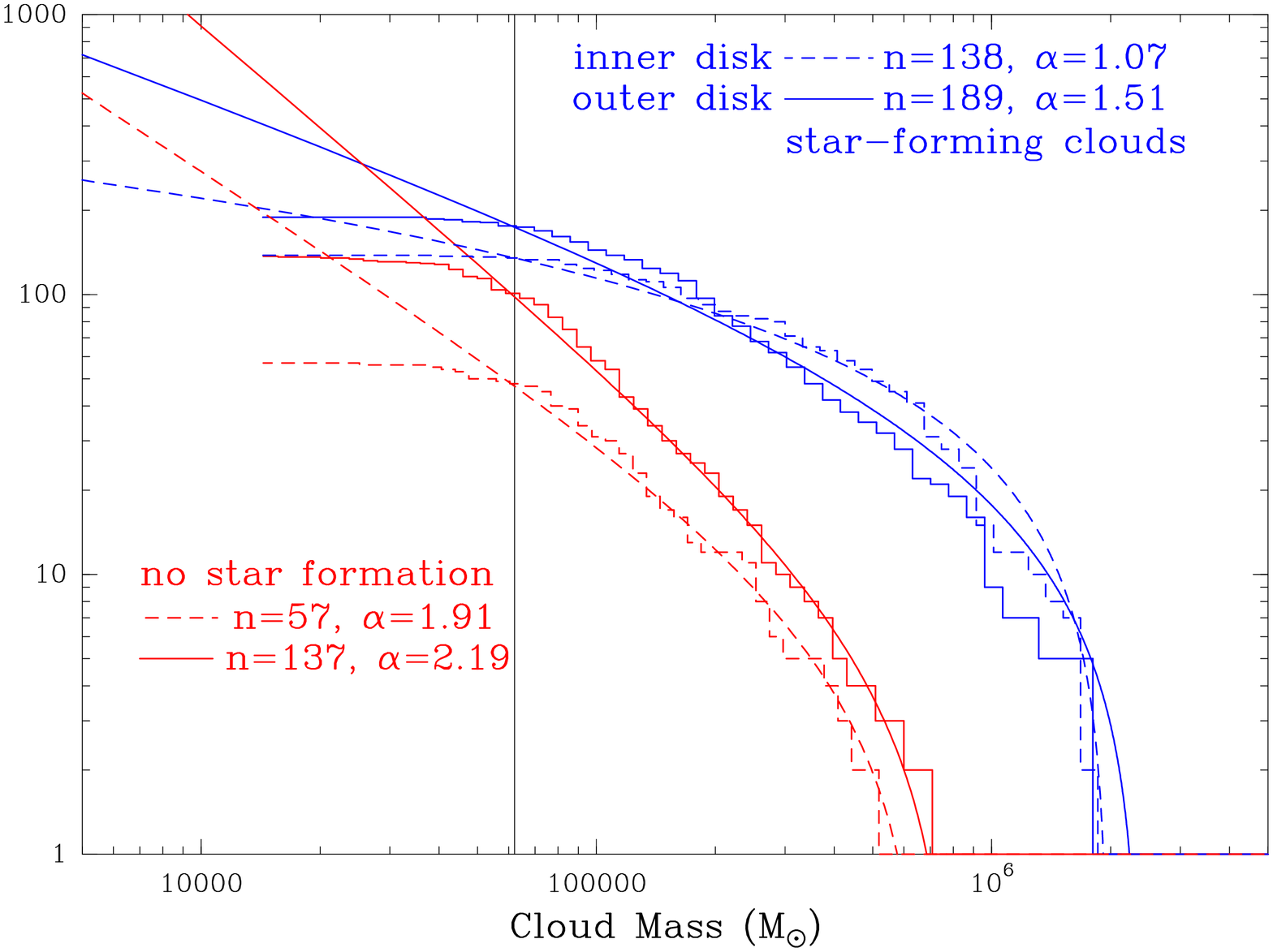}
	\caption{ Similar to Fig.~\ref{mass_spec} but separating inner and outer disk A and C clouds.  It is immediately apparent that galactocentric distance is less important than star formation.  The number of clouds and the slope of the mass function are given for each group.  See text for further details.}
	\label{mass_spec2} 
\end{figure}

The sample is large enough to be divided further in order to 
determine whether the steepening is primarily due to lack of star formation or position in disk.
We thus selected A and C clouds beyond 2 kpc from the center of M33, obtaining 98 A clouds above the completeness limit with an average galactocentric distance of 3.5 kpc and respectively 174 clouds and a distance of 3.9 kpc for the C clouds.
These are plotted as solid lines in Fig~\ref{mass_spec2}.  Despite the higher average galactocentric distance, the outer disk C clouds have significantly higher masses and a shallower mass spectrum than the A clouds.  In the figure, the number of clouds given is the total sample population, not the number above the completeness limit (given above).

Although A clouds are not common in the inner disk, we selected A and C clouds within respectively 2.5 and 2.8 kpc from the center, yielding 47 and 135 clouds above the completeness limit with an average galactocentric distance of 1.6 kpc for both samples.
Again, the mass spectra are very different, with the non-star-forming clouds being very similar to the outer disk population even when they belong to the inner disk.  {\it Clouds, and their mass spectra, change as star formation develops and progresses.}  This change is more important than their position in the disk although the change in CO luminosity is greater as a function of galactocentric distance than star-formation class.

\subsection{Dust temperatures with and without star formation}

\begin{figure}
	\centering
	\includegraphics[width=\hsize{}]{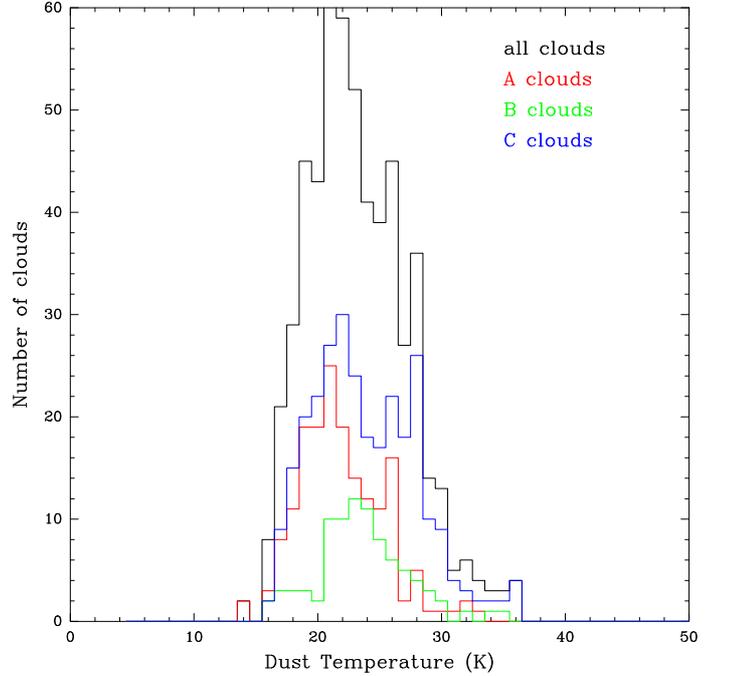}
	\caption{Representative dust temperatures for the big grain component in M33.  The decomposition into subsamples in terms of their star-forming properties shows that non-star-forming clouds have lower dust temperatures, although the bands used to estimate dust temperatures are completely independent of those used to classify the star formation in the clouds. }
	\label{herschel_fluxes} 
\end{figure}

Herschel SPIRE data are available for M33 and, following \citet{Braine10b}, we use the 250$\mu$m to 350$\mu$m flux ratio (after convolving the 250$\mu$m maps to the 350$\mu$m resolution) to estimate dust temperatures for the clouds.  We use a dust emissivity $\beta = 1.8$ for M33, following \citet{Tabatabaei14} for inner disk clouds, in order to calculate the temperatures.  While the absolute temperatures determined will vary,
 the fact that star-forming clouds have demonstrably higher dust temperatures (see Figure~\ref{herschel_fluxes}) than non-star-forming clouds does not change with the value of $\beta$ or wavelengths used.  

The non-star-forming ('A') clouds have significantly lower dust temperatures and FIR fluxes than the star-forming clouds, although  the fluxes have a very broad distribution for all cloud types.  The 'B' clouds, with embedded star formation but little or no H$\alpha$ emission, are presumably on average younger than those ('C') with exposed star formation.  The dust temperatures in the Herschel bands are not distinguishable between 'B' and 'C' clouds but the 250$\mu$m fluxes are slightly lower for the 'B' clouds.

\section{Rotation of molecular clouds}

As in all previous work on the subject (see references in introduction), we assume that velocity gradients reflect cloud rotation.
Rotation clearly results in velocity gradients.  However, other processes such as turbulence may be able to create velocity gradients \citep[see e.g.][for small scales]{Burkert2000}.  The velocity gradient is our only measurable and the systematic aspect of the gradients (Sect. 3.2) argues against other processes.
As in earlier work \citep[e.g.][]{Rosolowsky05}, the first moment of each spatial pixel in the cloud was calculated.  In our implementation, we used 5 velocity channels (13 $\kms$), centered on the central velocity of the cloud, to measure the first moment 
\begin{equation}
v_{(x,y)} = \sum_{i=cen-2}^{i=cen+2} v_i T_i dv / \sum_{i=cen-2}^{i=cen+2} T_i dv .  
\end{equation}
This was found to avoid bringing in too much noise (i.e. as when more channels are used) while still  covering the velocities occupied by the cloud.

This yields a velocity for each position in the cloud.  A plane is then fit to these velocities 
\begin{equation}
 v_{(x,y)} = ax + by + c
\end{equation}
where $a=\frac{\partial v}{\partial x}=\frac{\partial v}{\partial RA}$ and $b=\frac{\partial v}{\partial y}=\frac{\partial v}{\partial Dec}$ because $x$ and $y$ are the pixel numbers following the RA and Dec directions.  
 
The process is illustrated in Figure 1, which shows the CO(2--1) map of M33 with the cloud contours superposed and a zoom on a cloud showing the velocities as measured using Eq (1) in the top right panel, the fit to the velocities using Eq (2), and to the bottom right the residual ($v - v_{\rm fit}$).  This cloud (\#4 in the catalog) was chosen to illustrate the process for its rather clear gradient despite its elongated form.  It is worth noting that there is emission which CPROPS was not able to decompose into clouds and that the fraction of the CO emission not decomposed into clouds increases with radius.

\begin{figure}
	\centering
	\includegraphics[width=\hsize{}]{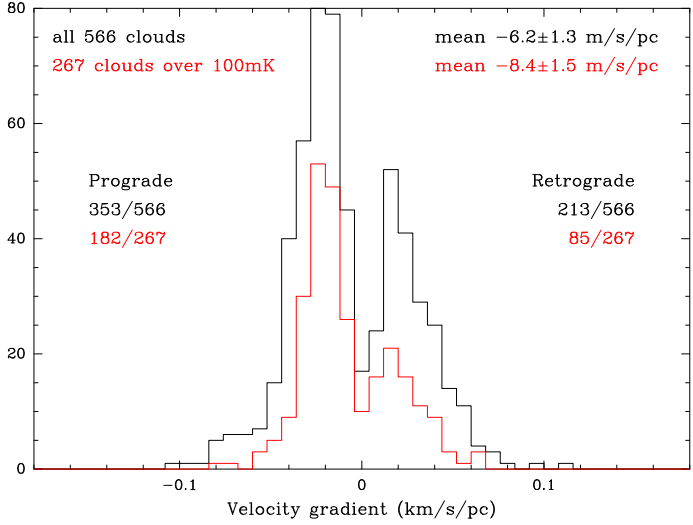}
	\caption{Histogram of cloud velocity gradients -- for the entire sample in black and for only the stronger clouds in red, where the line temperature averaged over the whole cloud is over 100mK T$_a^*$.  Prograde rotation is given a negative sign here because the galaxy rotation velocity increases with decreasing declination.  }
	\label{rot_fig2} 
\end{figure}

M33 is an inclined spiral whose North(-east)ern side is approaching, i.e. a higher negative velocity.  The near side of the disk is the western side.  The geometry is most easily visualized if one initially thinks of M33 as oriented N-S.  Thus, prograde follows the rotation of the disk, in which velocities become more negative to the North.  Rotating M33 counter-clockwise by 22.5$^\circ$ to its true orientation on the sky, things change little in that northern parts of a cloud have increasingly negative velocities when cloud rotation is prograde.  The gradients are measured as 
$\|   \bigtriangledown V   \| = \sqrt{(\frac{\partial v}{\partial RA})^2 + (\frac{\partial v}{\partial Dec})^2 }$ where $\frac{\partial v}{\partial RA}$ and $\frac{\partial v}{\partial Dec}$ come from the fits of a plane to the cloud velocities.  The sign of the gradient (negative for prograde, positive for retrograde in order to fit the true orientation of M~33) follows the sign of  
$\frac{\partial v}{\partial RA} sin(22.5)+\frac{\partial v}{\partial Dec} cos(22.5)$.

Figure \ref{rot_fig2} shows the cloud velocity gradients -- for the entire sample in black and for only the high signal-to-noise (S/N) clouds in red.
The fact that the distribution of velocity gradients extends to higher (absolute) values for the lower luminosity clouds is an immediate suggestion that noise is contributing to the observed gradients.  Moving to the more luminous clouds (red), not only is the distribution of velocity gradients narrower but it is more skewed towards negative (prograde) values.  

It may be worth noting that no difference in linewidth or in velocity gradient between star-forming (C clouds, cf. Sections 2.4 and 2.5) and non-star-forming clouds is found in our sample. 

\subsection{Beam smearing}

Molecular clouds are comparable in size to the angular resolution of our observations.  A typical (deconvolved) cloud radius is about 45 pc (Fig.~\ref{size-linewid}), to be compared with the 24pc beam half-power radius.  The goal of this section is to create mock clouds and compare the gradients derived before and after convolving with the telescope beam.  Intuitively, one expects gradients to  weaken as the angular resolution is degraded.

We create mock clouds with exactly the same gradients and sizes as those we have measured.  The gradients are injected as being linear and along the direction of the observed gradient.  The peak line temperature decreases linearly with distance to the cloud center until it reaches the noise level.  The lines are assumed gaussian.  This process is described in more detail in Section 4. No noise is added in this step as we wish to measure the effect of resolution and not noise (done later).

\begin{figure}
	\centering
	\includegraphics[width=\hsize{}]{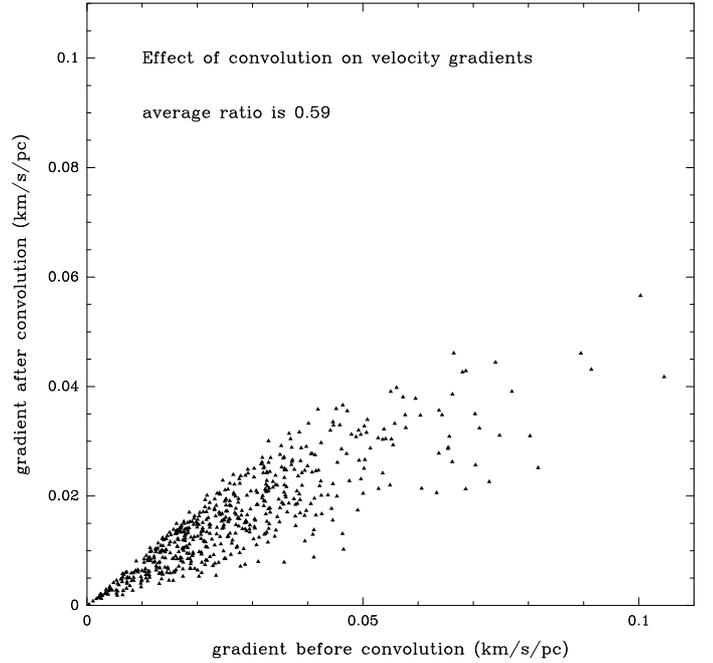}
	\caption{Comparison of the gradients measured pre- and post-convolution with the telescope beam. 
	On average, a correction factor of 1/0.59 should be applied to the observed gradients.}
	\label{conv} 
\end{figure}

After creating 566 mock clouds, we convolve them with a gaussian beam of half-power width $12\arcsec$.  We then follow exactly the same procedure to measure the gradients.  Fig.~\ref{conv} shows the comparison of the injected and recovered gradients (after convolution).  As could be expected, the post-convolution gradients are smaller, with an average ratio of 0.59.  We use this to correct the observed gradients in order to estimate rotation periods, angular momenta, and rotational energies.  None of the gradients after convolution are higher.

\subsection{Comparison with simulations}

Simulations appear to favor prograde rotation but the comparison with our measures is not always straightforward.  
\citet{Dobbs08} and \citet{Li17} 
use the specific angular momentum, defined as the angular momentum per unit mass averaged over the cloud, 
instead of the gradient.  The units are km s$^{-1}$ pc.

The angular momentum of a rotating disk with a surface density declining as $\Sigma(r) = \Sigma(r_0) (r/r_0)^{-1}$ is 
$L=\int_0^R r^2 \, \Omega \, \Sigma(r) \, 2\, \pi\, dr$.
Dividing by the mass of the disk ($2 \pi \Sigma_0 r_0 R$) yields a specific angular momentum of $j=\Omega R^2 / 3$ 
where $\Omega$ is the angular velocity and R the outer radius of the disk.  
A finite sphere with a density decreasing as $r^{-2}$ yields a similar result. 
If the disk has a constant surface density, then the specific angular momentum is $j=\Omega R^2 / 2$ \citep{Blitz93} but 
this is less likely.  
We thus consider the specific angular momentum of our clouds to be $j=(\Omega / 0.59) \, R^2 \, / 3$ where the 0.59 corrects for the underestimate in the velocity gradient due to beam smearing.

\begin{figure}
	\centering
	\includegraphics[width=\hsize{}]{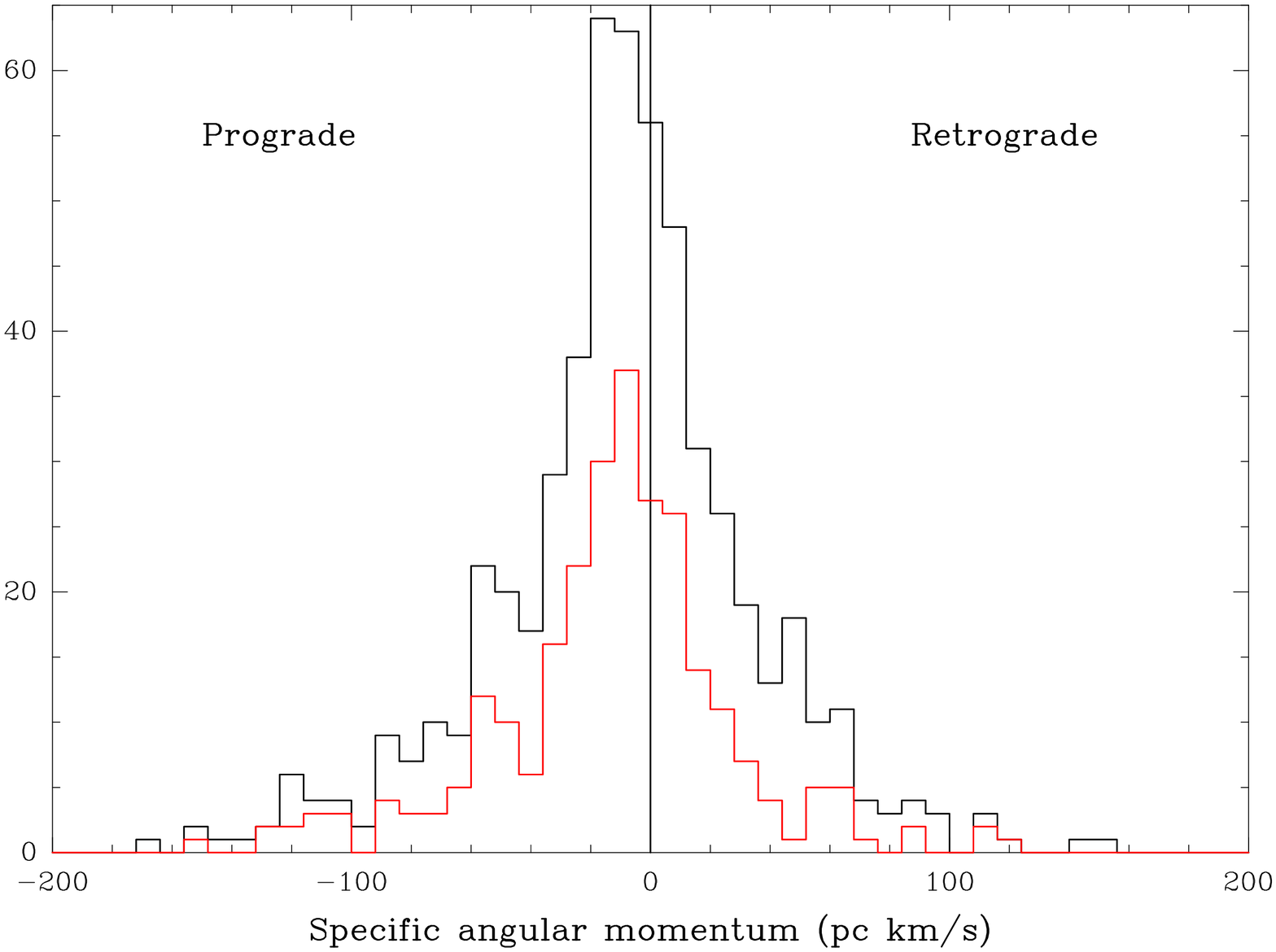}
	\includegraphics[width=\hsize{}]{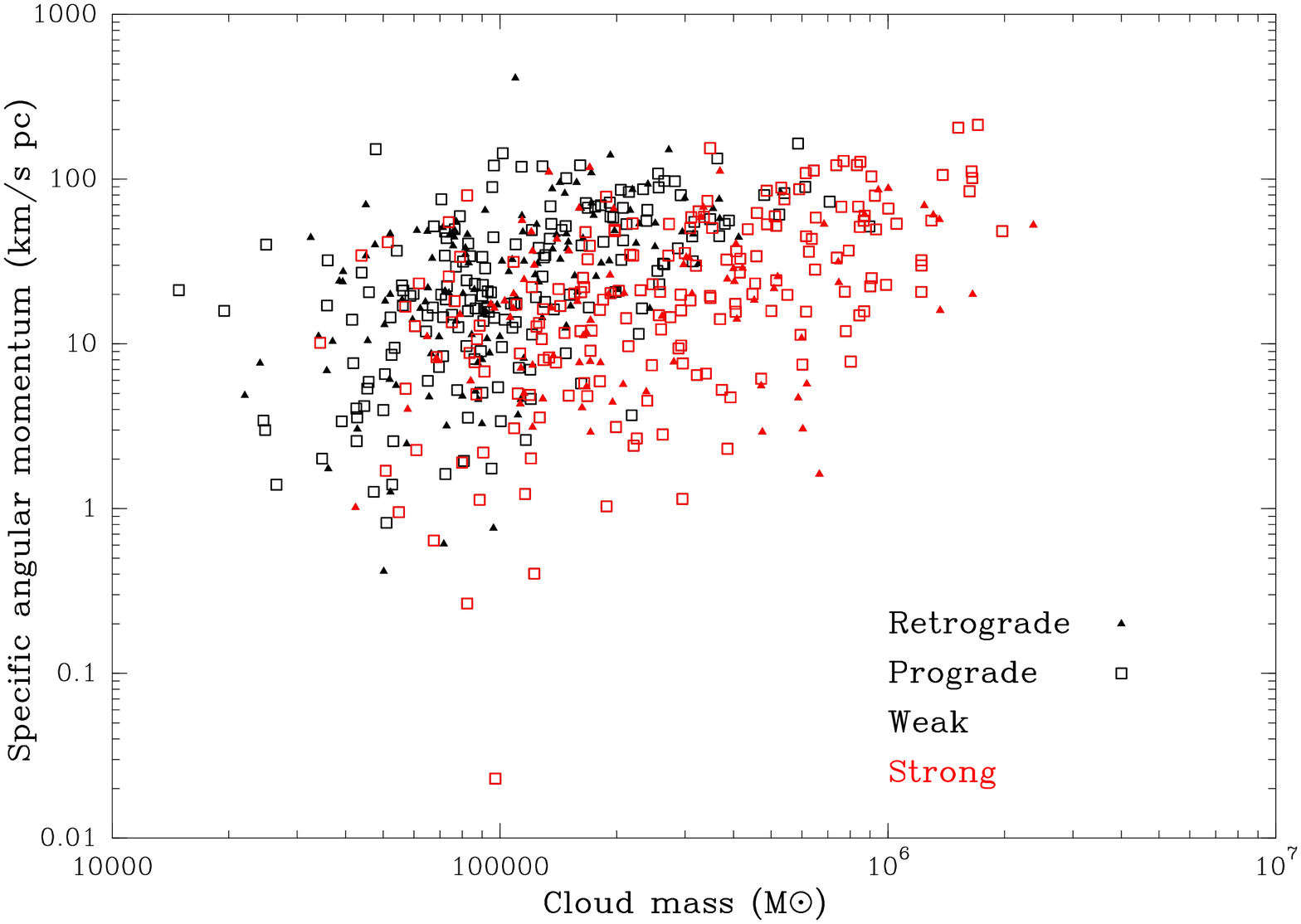}
	\caption{ Specific angular momenta of the M33 clouds.  Red lines or symbols indicate data for the CO-strong clouds. The top panel shows clearly that the asymmetry favoring prograde gradients (Fig.~\ref{rot_fig2}) is also present for the angular momenta.  The lower panel shows that the variation of angular momentum with cloud mass is very weak and was designed for comparison with  simulations.  }
	\label{ang_mom} 
\end{figure}

Fig.~\ref{ang_mom} shows the distribution of the angular momenta of our clouds.  The top panel is a histogram similar to the velocity gradients.  The bottom panel is intended to be useful for comparison with Figure 8 of \citet{Dobbs08}, giving the angular momenta as a function of cloud mass.  In both panels, the strong clouds are in red and retro/prograde rotation are separated either by sign (top) or by symbol (bottom).  Prograde rotation clearly dominates, as in the simulations where the self-gravity plays a role in the \citet{Dobbs08} simulations.  Comparing with \citet{Li17} Run VI, the distribution of the angular momenta in the M33 clouds appears narrower.  The \citet{Li17}  simulations have lower cloud masses and angular momenta for the high resolution Run VI, although the resolution of our observations is worse (suggesting that the difference would be greater if the spatial resolutions were closer).
Without more measurements of angular momenta of clouds, including in higher surface density galaxies than M33, it is difficult to be conclusive about the comparison with simulations.  

\section{Tests of velocity gradients}

We first need to convince ourselves that we measure real gradients.  The fact that the pro/retrograde differences are more pronounced for the high S/N sample is certainly a sign that this is the case.  In the following subsections, we examine the effect of beam smearing due to the resolution of our observations and create mock clouds with properties very similar to the real clouds in order to test our ability to retrieve cloud rotation in the presence of noise. 

The mock clouds are created using the masks of the real clouds, such that the sizes and shapes are perfectly represented.
The noise level of the cube is about 20 mK per channel \citep{Druard14} and we generate mock clouds with peak CO line temperatures T$_{max} = 100, 200, 400, 800$ mK in order to obtain varying S/N levels.
The pixels are then given a temperature 
\begin{equation}
T_{xyj} = [20+(T_{max}-20) \times (1-\frac{R}{R_{max}})] \, exp(-\frac{(v_j-v_{xy})^2}{2\Delta V^2}) 
\end{equation}
where x, y, and j represent respectively the pixel numbers along the RA, Dec, and velocity axes, and $R$ the distance from the center of the cloud (the position of the center of the cloud is returned by CPROPS).
The gradient is injected through the function relating $v_{xy}$ and the position through Eq. (2).
Thus, the line is centered on the velocity $v_{xy}$ and follows a gaussian with dispersion $\Delta V$.
As can be seen in Fig.~\ref{dist-dvco}, the linewidth at half power is roughly 7 $\kms$.  
We therefore inject a velocity dispersion $\Delta V = 3$ $\kms$, corresponding to a half-power linewidth of 7.05 $\kms$.
The central temperature decreases linearly with from T$_{max}$ to the noise level.
The sampling in space and velocity is the same as for the real cube (3\arcsec, 2.6$\kms$).

The next step is to add noise.  We add random gaussian noise using the well-tested $noise$ random number generator
within {\sc gildas}.  However, the data cube has undergone many transformations and the true noise may not be precisely gaussian.
Thus, we extract contiguous channels from signal-free regions of the cube which we use as noise.  In fact, as any transformations were designed to preserve particularly the region where signal is present, this is a sort of worst-case noise.

\subsection{No velocity gradient with noise}

\begin{figure}
	\centering
	\includegraphics[width=\hsize{}]{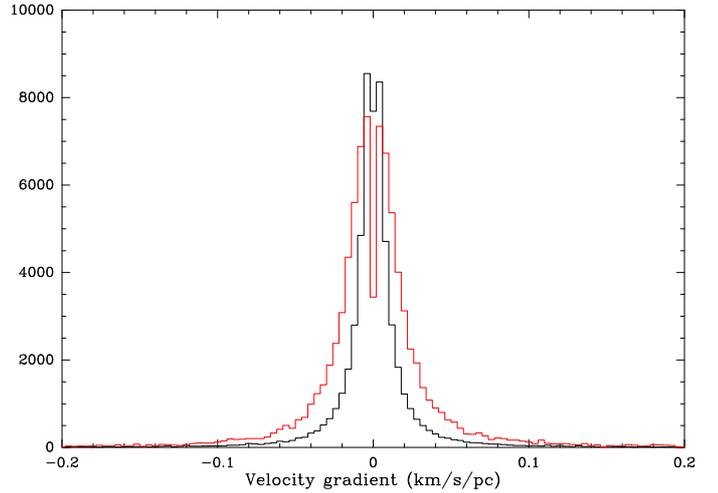}
	\caption{Histogram of gradients from noise. The black line shows the distribution of the gradients for purely random noise.  The red line shows the distribution when noise is from signal-free regions of the cube.  No velocity gradient has been injected here.}
	\label{zerograd} 
\end{figure}

We first examine what we obtain from mock clouds with no velocity gradient, {\it i.e.} with $v_{xy}$ constant.
Fig.~\ref{zerograd} shows the results for T$_{max}=200$ mK using both purely random noise and noise taken from signal-free but unoptimized regions of the cube.  The gradients, for equivalent clouds and noise levels, are clearly more dispersed with the "real" cube noise.  The dispersions, as measured by the full width at half power of the distrbution (divided by 2.35 to give the equivalent for a gaussian), are approximately 0.008 and 0.015 $\kms {\rm pc}^{-1}$for the random and cube noise injection.  T$_{max}=200$ mK corresponds roughly to the median signal in the cloud sample.  

We have also tested with higher and lower S/N levels.  For T$_{max}=100$ mK, the dispersion increases to 0.014 and 0.029 $\kms {\rm pc}^{-1}$for respectively random and cube noise.  For T$_{max}=400$ mK, the dispersion decreases to 0.005 and 0.0085 $\kms {\rm pc}^{-1}$for respectively random and cube noise.  

\section{Evaluating uncertainties on cloud rotation}

\begin{figure}
	\centering
	\includegraphics[width=\hsize{}]{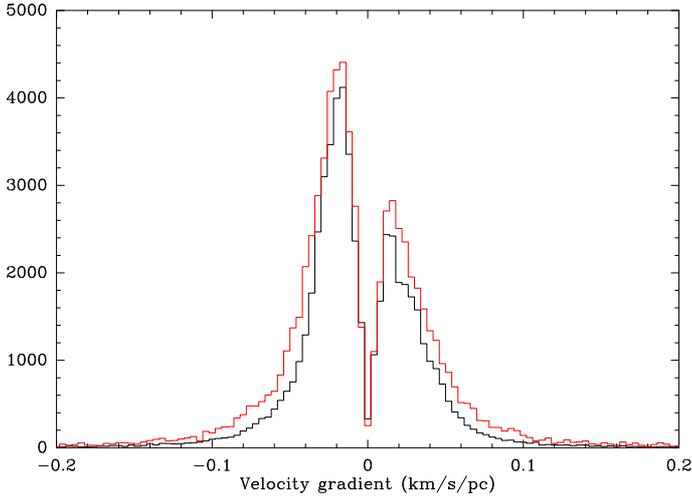}
	\caption{ Like Fig.~\ref{zerograd} except that the observed gradients (see Fig.~\ref{rot_fig2}) have been injected, noise added, and the gradients recovered.  Black line shows the distribution of the gradients for purely random noise.  The red line shows the distribution when noise from signal-free regions of the cube.  As in Fig.~\ref{zerograd}, T$_{max}=200$ mK.}
	\label{grad200} 
\end{figure}

The same operations were done with the observed gradients, creating the same clouds but injecting the velocity gradient deduced from the observations for each cloud.  Should there be a link between size or shape and the velocity gradient deduced from calculating the first moment, the link would be preserved in these tests.  Fig.~\ref{grad200} shows the distribution of the retrieved gradients.  The clouds were created with the observed gradients and shapes and then noise was added and the gradients remeasured.  This can be directly compared with Fig.~\ref{rot_fig2}.  The distribution in Fig.~\ref{grad200} is of course wider because the gradients from Fig.~\ref{rot_fig2} have been injected and then noise added.  The process was repeated with T$_{max}=100$ mK and the distribution is significantly wider but the prograde-retrograde asymmetry is nonetheless preserved.
We are thus confident that the gradients retrieved reflect the true distribution of velocity gradients, although the distribution is likely broadened by the presence of the noise in the datacube.

\subsection{Galactic gradient}

M~33 itself has a velocity gradient due to rotation.  Since the position angle of M33 is close to vertical, there is little gradient expected along the RA ($x$) axis but there is a negative gradient along the Dec ($y$) axis because at higher Declinations the velocity is more negative.  We calculate this for axisymmetric rotation assuming the rotation curve given in Eq. 18 of  \citet{Lopez17}:
$$ V(r) = V_0 \frac{(r/r_0) +d}{(r/r_0)+1} $$
where $V_0 = 139.2$ $\kms$, $r_0=1.3$ kpc, and $d=0.12$.

\begin{figure*}
	\centering
	\includegraphics[width=\hsize{}]{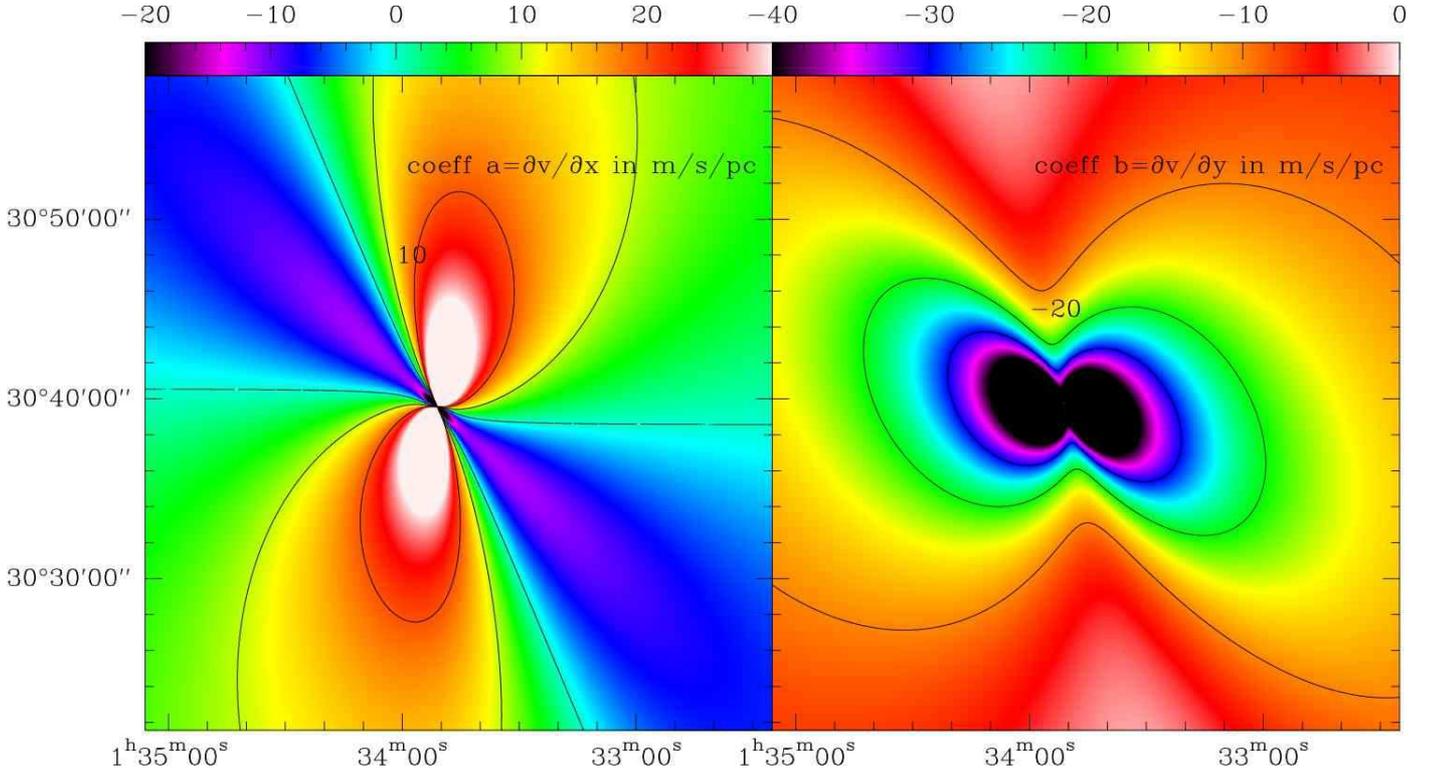}
	\caption{Coefficients $\frac{\partial v}{\partial RA}$ (left) and $\frac{\partial v}{\partial Dec}$ (right) calculated for adjacent pixels  using the rotation curve above.  Units are meter~s$^{-1}$~pc$^{-1}$ and the color wedges are shown at the top of each panel. In the left panel, contours are drawn at 0, 10, and 20 m~s$^{-1}$~pc$^{-1}$ and in the right panel at -10, -20, and -30 m~s$^{-1}$~pc$^{-1}$. }
	\label{rot_fig3} 
\end{figure*}

Figure \ref{rot_fig3} shows the local velocity gradients $a=\frac{\partial v}{\partial RA}$ and $b=\frac{\partial v}{\partial Dec}$ derived from the axisymmetric rotation curve.  The plot of $\frac{\partial v}{\partial Dec}$ is negative everywhere with fairly high (absolute) values  but $\frac{\partial v}{\partial RA}$ has both negative and positive regions with a positive average.  
These values come exclusively from the rotation curve and thus include differential rotation and thus shear.

Figure \ref{rot_fig3} is not very intuitive.  In order to qualitatively understand the negative and positive zones, let us think of isovelocity curves of a differentially rotating spiral disk with a monotonically increasing rotation curve (the so-called spider diagram).  When the major axis is vertical (N--S), then the only horizontal iso-velocity curve (i.e. $\frac{\partial v}{\partial RA}=0$) is along the minor axis. There are no vertical (i.e. $\frac{\partial v}{\partial Dec}=0$) iso-velocity curves, such that for velocities decreasing towards the North, $\frac{\partial v}{\partial Dec} < 0$ everywhere.  Now let us rotate the diagram counterclockwise slightly.  In the northern half, we will have a locus of $\frac{\partial v}{\partial Dec}=0$ points just to the left of the major axis, where the iso-velocity curves are briefly horizontal.  Slightly above the minor axis and to the left, the isovelocity curve which went slowly upwards pre-rotation now is approximately flat, leading to another series of $\frac{\partial v}{\partial Dec}=0$ points.  The same is true by symmetry to the South.  No such region where $\frac{\partial v}{\partial Dec}=0$ is present to the upper right or lower left.  The magnitude can be understood by imagining how closely spaced (along RA or along Dec) the isovelocity curves are (for equal velocity spacing).  This is why the highly negative regions of $\frac{\partial v}{\partial Dec}$ are close to the minor axis.  Similarly, $\frac{\partial v}{\partial RA}$ is high where isovelocity curves are closely spaced and close to vertical.  In all cases, the velocity gradients due to galactic rotation are larger near the center where the rotation rises sharply.

\subsection{Comparison of cloud and galactic gradients}

\begin{figure*}
	\centering
	\includegraphics[width=\hsize{}]{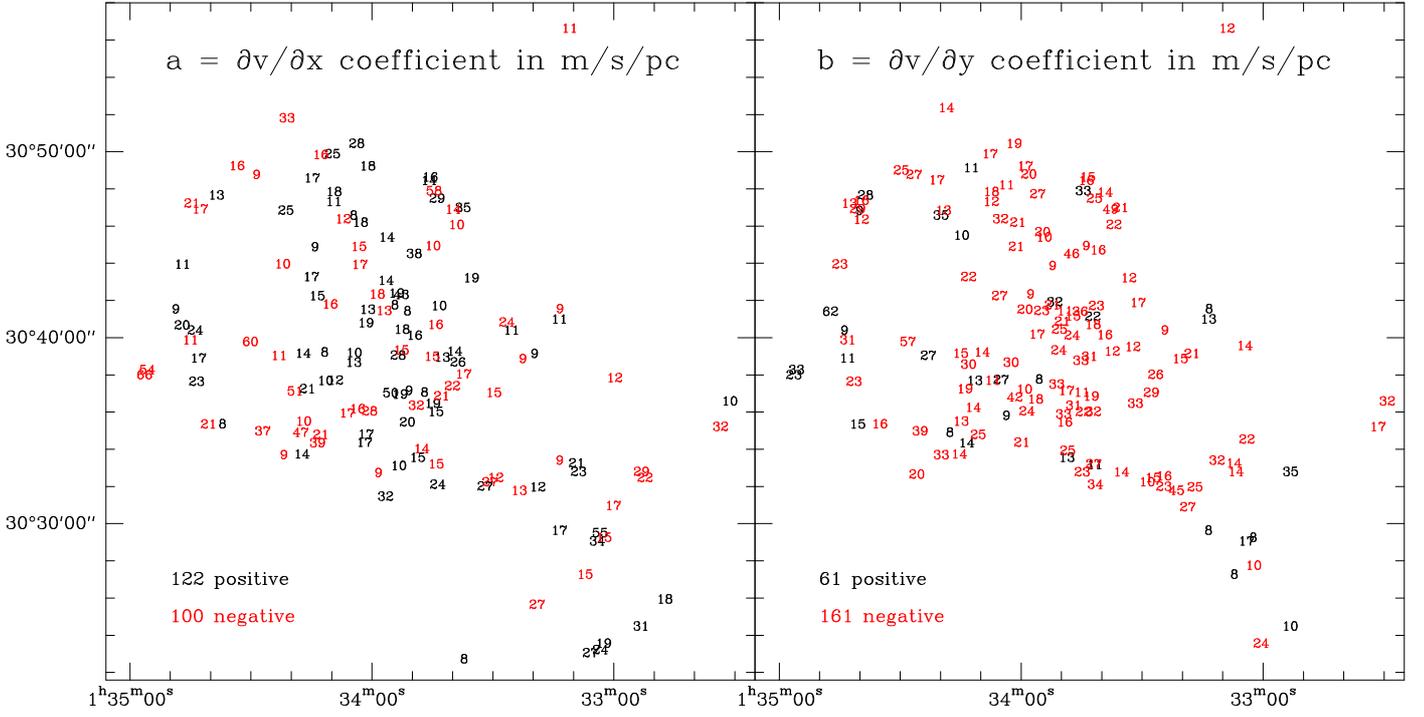}
	\caption{Coefficients $a=\frac{\partial v}{\partial RA}$ and $b=\frac{\partial v}{\partial Dec}$ for the 222 stronger clouds ($T_{CO} > 0.11$K), shown only when above 8 m~s$^{-1}$~pc$^{-1}$ in absolute value in order to reduce the influence of noise and make the numbers legible. Negative values are shown in red and positive in black. The numbers of negative and positive values are given in the panels.  Because the panels only show the higher (absolute) values, the disproportion in the right panel is actually greater: 31 values $>8$ and 116 values $< -8$. }
	\label{rot_fig4} 
\end{figure*}

Having fit a plane to the velocities of the pixels making up each cloud, we have the gradients along the RA and Dec axes and we can look for patterns.  Given Fig.~\ref{rot_fig2} which shows that we can have more confidence in the high-luminosity clouds, we plot the gradients for the stronger clouds in Fig. \ref{rot_fig4} in a way that can be compared directly with Fig.~\ref{rot_fig3}.  Let us consider the "null" hypothesis to be that clouds {\it on average} are not rotating with respect to their surroundings, {\it i.e.} they rotate with the galaxy.  
Our results are close to this "null" hypothesis (cf. Figures ~\ref{rot_fig3} and \ref{rot_fig4}).
If correct, an implication is that the cloud formation mechanism has little influence on the velocity gradient.  

From Fig.~\ref{rot_fig2}, a typical prograde rotation velocity is $\la 0.03$ $\kms$~pc$^{-1}$.  Fig. \ref{rot_fig4} shows that this is a good representative value for the CO-strong clouds.   Including a factor $1/0.59$ to compensate for the beam smearing discussed earlier, this yields $\Omega \la 0.05$ $\kms$~pc$^{-1}$.  The rotation period is thus about $T=\frac{2 \pi}{\Omega} \approx 120$ Myr, which is similar to the rotation period of the inner disk of M33.

Thus, not only are real gradients measured in these clouds but we are able to show that they are dominated by prograde rotation despite the extremely low values.  The rotation periods are longer than cloud lifetimes and comparable to the Galactic rotation period.  The link between Fig.~\ref{rot_fig3} and Fig.~\ref{rot_fig4} is real: the average observed $\frac{\partial v}{\partial RA} = 2.7\pm1.2$ m~s$^{-1}$~pc$^{-1}$ where $\frac{\partial v}{\partial RA} > 0$ in Fig.~\ref{rot_fig3} but the average observed $\frac{\partial v}{\partial RA} = -2.0\pm1.5$ m~s$^{-1}$~pc$^{-1}$ where $\frac{\partial v}{\partial RA} < 0$ in Fig.~\ref{rot_fig3} and all averages are negative for $\frac{\partial v}{\partial Dec}$. 

\subsection{Magnitude of velocity gradients}

Is rotation a significant hindrance to cloud collapse?
Adopting $\Omega \approx 0.05$ $\kms {\rm pc}^{-1}$as typical of a "rotating" cloud, we can compare the rotational kinetic energy with the gravitational potential energy or the edge velocities with escape velocities.  
Adopting $M=2 \times 10^5$ \msun and $R=30$ pc as representative values, the rotational kinetic energy is $E_{rot} \approx 10^{48}$ ergs whereas the gravitational potential energy is nearly $E_{grav} \approx 10^{50}$ ergs.  Similarly, the rotation velocity at the cloud edge could be expected to be $v \approx 1.5\kms$ but the escape velocity is much higher, $v_{esc} \approx 7.5$ $\kms$.  This large difference shows also that rotation contributes little to the overall support and line width of the cloud.


%

\section{Conclusions}

In this work, we have shown, for the first time to our knowledge, that molecular clouds rotate and that their rotation is very slow but measurable from our high-quality data.  This relies on the assumption, as in previous work, that rotation can be deduced from velocity gradients. The rotation tends to be prograde.  The majority of molecular clouds have an angular velocity below .03 $\kms {\rm pc}^{-1}$(.05 $\kms {\rm pc}^{-1}$after correcting for beam-smearing), yielding a rotation period greatly superior to the cloud lifetimes of about 15 Myr in M~33 \citep{Corbelli17}.  The rotation contributes (very) little to the support of the cloud against gravity.  Simulations as well as classical calculations \citep[e.g.][]{Rosolowsky03} tend to find higher angular velocities.
At (much) smaller scales rotation is clearly present: stars and proto-stars have disks (which rotate) and rotation was also observed in the massive proto-stellar core W43-mm1 \citep{Jacq16}. 

Not only do molecular cloud mass spectra steepen with galactocentric distance, but the mass spectrum appears to depend even more strongly on whether the clouds host active star formation.  At equivalent galactocentric distance, molecular clouds which form stars have considerably flatter mass spectra than those without star formation.

Comparing the molecular clouds in M~33 with those in other nearby galaxies, a displacement in the size-linewidth relation appears in that lower metallicity systems have narrower CO lines for comparable cloud size.  There is also a trend for cloud linewidths to become narrower with increasing galactocentric distance.  Some degeneracy is present in these measurements as both metallicity and stellar surface density decrease with galactocentric distance and subsolar metallicity galaxies tend to have lower stellar surface densities.

\begin{acknowledgements}
The authors would like to gratefully acknowledge several students from the University of Bordeaux who did projects on various aspects of the data: Jimmy Mata, Marc-Robert Antoine, Thomas Goncalves, and Cyril Lenain.  
\end{acknowledgements}

\bibliographystyle{aa}
\bibliography{jb}

\end{document}